# Anomalous broadening of specific heat jump at $T_c$ in high-entropy-alloy-type superconductor $Tr$Zr$_2$

Md. Riad Kasem[1], Aichi Yamashita[1]*, Taishi Hatano[2], Yosuke Goto[1], Osuke Miura[2]

Yoshikazu Mizuguchi[1]

[1] Department of Physics, Tokyo Metropolitan University, Hachioji, Japan
[2] Department of Electrical Engineering and Computer Science, Tokyo Metropolitan University, Hachioji, Japan

E-mail: aichi@tmu.ac.jp

**Abstract**

A high-entropy-alloy-type (HEA-type) superconductor is new category of highly disordered superconductors. Therefore, finding brand-new superconducting characteristics in the HEA-type superconductors would open new avenue to investigate the relationship between structural disorder and superconductivity. Here, we report on the remarkable broadening of specific heat jump near a superconducting transition tempreature ($T_c$) in transition-metal zirconides ($Tr$Zr$_2$) with different mixing entropy ($\Delta S_{mix}$) at the $Tr$ site. With increasing $\Delta S_{mix}$, the superconducting transition seen in specific heat became broader, whereas those seen in magnetization were commonly sharp. Therefore the broadening of specific heat jump would be related to the microscopic inhomogeneity of the formation of Cooper pairs behind the emergence of bulk superconductivity states.

Keywords: high-entropy-type superconductor, transition-metal zirconide, specific heat

## 1. Introduction

High-entropy alloys (HEAs) are defined as alloys composed of five or more constituent elements with a concentration range between 5 to 35 at%. Due to their high performance as high-temperature materials, structural materials, medical materials, etc., originating from high configurational mixing entropy ($\Delta S_{mix}$) defined as $\Delta S_{mix} = -R \Sigma_i c_i \ln c_i$, where $c_i$ and $R$ are compositional ratio and the gas constant, HEAs have been extensively studied in the field of material science in recent years [1–3]. In the field of physics and chemistry, the discovery of superconductivity in a HEA Ti-Zr-Hf-Nb-Ta triggered exploration of new HEA superconductors [4–13]. In addition, the robustness of superconductivity in a HEA under pressures up to 190 GPa was reported [14]. Although the HEA concept has expanded the range of exploration of alloy superconductors, the relationship between superocnductivity and the effects of $\Delta S_{mix}$ have not been deeply studied, so far. Therefore, further development of HEA-type superconductors, not only simple alloys but also HEA-type compounds with two or more crystallographic sites, had been desired [15].

In 2018, we reported the synthesis and superconducting properties of HEA-type layered BiS$_2$-based superconductor $RE$(O,F)BiS$_2$ with five different rare-earth elements ($RE$ = La, Ce, Pr, Nd, Sm) [16]. Notably, an improvement of superconducting properties due to the suppression of local structural disorder in the BiS$_2$ conducting layers were observed as $\Delta S_{mix}$ increased [17]. This fact suggests that the introduction of HEA-type site in compounds having more than two crystallographic sites modifies local crystal structure and tune physical properties. Furthermore, HEA-type layered compounds were recently synthesized in the series of cuprate and van der Waals superocnductors in both polycrystalline and single-crystal forms [18–20].

Another HEA-type superconducting compounds are rock-salt-type (cubic NaCl-type) metal tellurides ($M$Te) and chalcogenides ($MCh$) with $M$ = Ag, Cd, In, Sn, Sb, Pb, Bi [21–23]. The superconducting transition temperature ($T_c$) of HEA-type $M$Te was clearly lower than that of pure or low-entropy $M$Te superconductors [22], and the highest $T_c$ among the HEA-type $M$Te was about 2.6 K in (Ag,In,Sn,Pb,Bi)Te. The low $T_c$ could be related to the high disorder in $M$-Te bonds generated by the HEA-type $M$ site. That would directly affect electronic states, and hence superconductivity was suppressed. To obtain further knowledge about the relationship between superconductivity and the introduction of a HEA site in compounds, we recently developed tetragonal HEA-type transition-metal zirconides $Tr$Zr$_2$ ($Tr$ = Fe, Co, Ni, Cu, Rh, Ir) with a CuAl$_2$-type (space group: #140, $I4/mcm$) structure (see Fig. 1(c)) [24,25]. $Tr$Zr$_2$ has a relatively high $T_c$, which is a merit when deeply studying the essence of the HEA effects in compounds: $T_c$ of pure $Tr$Zr$_2$ are 0.17, 5.5, 1.6, 11.3, 7.5 K for $Tr$ = Fe, Co, Ni, Rh, and Ir, respectively [26]. In addition, physical properties of CoZr$_2$ single crystals have been expensively investigated [27], which suggested that electronic states originating from both $Tr$ and Zr atoms are essential to superconductivity. Therefore, in this study, we investigated the specific heat of $Tr$Zr$_2$ superconductors with different $\Delta S_{mix}$ (zero-, low-, and high-entropy compositions) to reveal the essence of the effect of the introduction of an HEA site in superconducting compounds. Here, we report on the remarkable broadening of specific heat jump at $T_c$ in samples with higher $\Delta S_{mix}$.

## 2. Experimental details

Polycrystalline samples of $Tr$Zr$_2$ ($Tr$ = Fe, Co, Ni, Cu, Rh, Ir) were prepared by arc melting in an Ar atmospher as described in Ref. 25. Powders of pure metals, Fe (99.9%), Co (99%), Ni (99.9%), Cu (99.9%), Rh (99.9%), and Ir (99.9%), were mixed with a certain composition and pelletized. The metal pellet and a plate of pure Zr (99.2%) were used as starting materials for arc melting. The arc-melting was repeated after turning over the sample three times to homogenize the sample. The phase purity and lattice constants were investigated by powder X-ray diffraction (XRD) by the $\theta$-$2\theta$ method using Cu-K$\alpha$ radiation on Miniflex600 equippted with a high resolusion detector D/tex-Ultra (RIGAKU). The lattice constants were calculated by refining the obtained XRD patterns using a Rietveld analysis software RIETAN-FP [28]. Crystal structure image was drawn using VESTA [29]. The chemical composition of the obtained samples were investigated using energy-dispersive X-ray spectroscopy (EDX) on TM3030 (Hitachi Hightech) using SwiftED analyzer (Oxford). The temperature ($T$) dependence of magnetization was measured using a superconducting intereference device (SQUID) on MPMS-3 (Quantum Design). The temprature dependence of specific heat was measured by a relaxation method on PPMS (Quantum Design) under magnetic fields ($\mu_0 H$) 0 and 9 T.

## 3. Results and discussion

We synthesized five $Tr$Zr$_2$ polycrystalline samples whose nominal compositions are CoZr$_2$ (#A), Co$_{0.3}$Ni$_{0.4}$Rh$_{0.3}$Zr$_2$ (#B), Fe$_{0.2}$Co$_{0.3}$Ni$_{0.2}$Rh$_{0.3}$Zr$_2$ (#C), Fe$_{0.2}$Co$_{0.2}$Ni$_{0.2}$Rh$_{0.2}$Ir$_{0.2}$Zr$_2$ (#D), and Fe$_{0.1}$Co$_{0.2}$Ni$_{0.2}$Cu$_{0.1}$Rh$_{0.2}$Ir$_{0.2}$Zr$_2$ (#E), in which one, three, four, five, and six transition metal elements, respectively, are solved in the $Tr$ site. Hereafter, we call samples with labels (#A–#E) or nominal compositions. The $Tr$ concentration was determined as a $T_c$ becomes close to 5 K because CoZr$_2$ (#A) has a $T_c$ was reported as ~5.5 K [26,27], so that we could discuss the effects of high-entropy alloying for the $Tr$ site on the sharpness of superconducting transitions. The obtained button samples were silver in colour.

Figure 1(a) shows the powder XRD patterns of all the samples. The peaks of the main phase can be indexed with a CuAl$_2$-type tetragonal model [24,25]. Although an impurity phase of cubic-type CoZr$_2$ (space group: #227), which is not a superconductor, was detected in Co$_{0.3}$Ni$_{0.4}$Rh$_{0.3}$Zr$_2$ (#B), no peak split was observed for all samples (see Fig. 1(b)), indicating the homogeneous solution of $Tr$ elements in the samples. Note that the EDX analyses also demonstrated no phase separation, particularly on the $Tr$-site composition, in those examined samples. The estimated lattice constants are listed in table I with the compositional and physical parameters.

Figures 2(a–e) show the temperature dependence of magnetization with zero field cooling (ZFC) and field cooling (FC). Large diamagnetic signals corresponding to the emergence of superconductivity are observed below $T_c^M$ = 6.0, 4.6, 5.3, 5.4 and 5.7 K for #A, #B, #C, #D, and #E, respectively. Previously, we have reported that $T_c$ shows a dependence on the lattice constant $c$ and/or $Tr$-Zr bonds along the $c$-axis [25]. It is possibly due to the tuning of the electronic density of states near the Fermi energy, which are composed by both Zr-$d$ and $Tr$-$d$ electrons. Here, we found that the $T_c$ of present samples also obeyed the trend in $Tr$Zr$_2$ (Supplemental data).

To confirm the bulk nature of observed superconductivity for all the samples, specific heat ($C$) measurements were performed. Figures 2(f–j) display the temperature dependence of $C$ under 0 and 9 T. A clear jump was observed for all the samples under 0 T, which indicates the bulk nature of superconductivity. The Sommerfeld coefficient ($\gamma$) and the coefficient for the lattice contribution ($\beta$) were estimated using the data obtained in non-superconducting states under 9 T. The estimated $\gamma$ were 22.20(4), 20.73(7), 22.93(8), 20.59(6) and 21.33(6)

mJ/mol·K$^2$, and $\beta$ were 0.332(5), 0.436(1), 0.428(1), 0.449(1) and 0.491(1) mJ/mol·K$^4$ for #A, #B, #C, #D, and #E, respectively. The $\beta$ slightly increased from 0.33 to 0.49 mJ/mol·K$^4$ with increasing $\Delta S_{mix}$. Consequently, the Debye temperature ($\Theta_D$) decreased with increasing $\Delta S_{mix}$ (see Table I). To clarify the characteristics of the superconducting jump, electronic contribution ($C_{el}$) was calculated by subtracting $\beta T^3$ from $C$, and plotted in figures 2(k–o). From the temperature dependences of $C_{el}/T$, the $\Delta C_{el}/\gamma T_c^C$ parameter was estimated as 1.16, 1.32, 1.13, 1.34, and 1.35 for #A, #B, #C, #D, and #E, respectively, where $T_c^C$ denotes a $T_c$ estimated from the $C$ analysis of entropy balance in the superconducting transition. The estimated $\Delta C_{sc}/\gamma T_c$ values are slightly smaller than $\Delta C_{sc}/\gamma T_c = 1.43$, which is expected from conventional weak-coupling pairing [30], indicating that the $Tr$Zr$_2$ samples would be weak-coupling superconductors [27]. Here, we unexpectedly observed the broadening of specific heat jump in HEA-type samples (#D and #E).

To compare the degree of the broadening of specific heat jump, the $C$ related to superconducting states ($C_{sc}$) was plotted against $T/T_c$ in Fig. 3. Notably, the degree of broadening of the jump systematically becomes broader with increasing $\Delta S_{mix}$, and a remarkable broadening of the specific heat jump was observed for HEA-type $Tr$Zr$_2$ (#D and #E). Although there are not enough number of reports on the specific heat measurement for HEA-type superconductor, similar broadening was also observed in HEA-type Nb$_3$Sn system [31] and (Co,Ni,Cu,Rh,Ir)Zr$_2$ [24]. Considering the sharp transitions in magnetization shown in Figs. 2(a–e), we assume that there are no large-scale phase separations with grains with different $T_c$. Hence, the broadening trend observed in $C$ is originating from microscopic background rather than bulk characteristics. Since the HEA-type materials possess highly dispersed $Tr$-Zr bonds, which should result in blurred band dispersion and the Fermi surfaces. In such condition of electronic states, the broadening of superconducting gap opening as a function of temperature would be explained. We expect that the superconducting characteristics observed in HEA-type superconducting compounds is related to multi-gap superconducting states, which was typically observed in MgB$_2$ and Fe-based superconductors [32,33]. To reveal the origins of the phenomena, further information from the experiments, which can directly observe the superconducting gap structure such as scanning tunneling microscopy (STM), and information regarding local structure and the bond-length dissipation, are desired.

## 4. Conclusion

We reported the synthesis and superconducting properties of CuAl$_2$-type $Tr$Zr$_2$ superconductors, CoZr$_2$, Co$_{0.3}$Ni$_{0.4}$Rh$_{0.3}$Zr$_2$, Fe$_{0.2}$Co$_{0.3}$Ni$_{0.2}$Rh$_{0.3}$Zr$_2$, Fe$_{0.2}$Co$_{0.2}$Ni$_{0.2}$Rh$_{0.2}$Ir$_{0.2}$Zr$_2$, and Fe$_{0.1}$Co$_{0.2}$Ni$_{0.2}$Cu$_{0.1}$Rh$_{0.2}$Ir$_{0.2}$Zr$_2$, with systematically tuned $\Delta S_{mix}$. Polycrystalline samples were prepared using pure metals by arc melting. The composition of the obtained samples were investigated by EDX and confirmed to be close to the nominal compositions. The $Tr$-site composition for Fe$_{0.2}$Co$_{0.2}$Ni$_{0.2}$Rh$_{0.2}$Ir$_{0.2}$Zr$_2$ and Fe$_{0.1}$Co$_{0.2}$Ni$_{0.2}$Cu$_{0.1}$Rh$_{0.2}$Ir$_{0.2}$Zr$_2$ satisfied with the typical definition of HEA. Superconducting transitions were observed for all the samples through magnetization and specific heat measurements. The bulk nature of superconductivity was confirmed from the specific heat jump at $T_c$. Although the low-entropy samples (#A and #B) exhibited a sharper specific heat jump, remarkable broadening was observed in middle-entropy (#C) and HEA-type (#D and #E) samples. Even though the origin is not clear at current stage, the present data indicates the modification of the superconducting gap opening by the HEA effect ,which is behind the bulk characteristics of superconductivity. The broadening phenomena of specific heat jump at $T_c$ in HEA-type superconducting compounds provide us with an additional pathway for investigations on the relationship between superconductivity and the effects of disorder in compounds.

## Acknowledgements

The authors thank T. D. Matsuda for his supports in experiments. This work was partly supported by Tokyo Metropolitan Government Advanced Research (H31-1) and JSPS-KAKENHI (18KK0076).

## Table

Table I. Compositional, structural, and specific heat parameters of $Tr\text{Zr}_2$ samples.

| Label | #A | #B | #C | #D | #E |
|---|---|---|---|---|---|
| $Tr$-nominal | Co | $Co_{0.3}Ni_{0.4}Rh_{0.3}$ | $Fe_{0.2}Co_{0.3}Ni_{0.2}Rh_{0.3}$ | $Fe_{0.2}Co_{0.2}Ni_{0.2}Rh_{0.2}Ir_{0.2}$ | $Fe_{0.1}Co_{0.2}Ni_{0.2}Cu_{0.1}Rh_{0.2}Ir_{0.2}$ |
| $Tr$-EDX | - | $Co_{0.25}Ni_{0.39}Rh_{0.36}$ | $Fe_{0.22}Co_{0.27}Ni_{0.18}Rh_{0.33}$ | $Fe_{0.22}Co_{0.22}Ni_{0.21}Rh_{0.21}Ir_{0.14}$ | $Fe_{0.09}Co_{0.20}Ni_{0.20}Cu_{0.17}Rh_{0.19}Ir_{0.15}$ |
| $\Delta S_{mix}/R$ | 0 | 1.08 | 1.36 | 1.60 | 1.76 |
| $a$ (Å) | 6.3597(7) | 6.4579(5) | 6.4375(3) | 6.4517(2) | 6.4995(4) |
| $c$ (Å) | 5.5094(6) | 5.4141(6) | 5.5017(3) | 5.5226(2) | 5.4918(4) |
| $T_c^M$ (K) | 6.0 | 4.6 | 5.3 | 5.4 | 5.7 |
| $T_c^C$ (K) | 5.5 | 4.3 | 4.9 | 4.7 | 5.0 |
| $\gamma$ (mJ/K$^2$mol) | 22.20(4) | 20.73(7) | 22.93(8) | 20.59(6) | 21.33(6) |
| $\Theta_D$ (K) | 260 | 237 | 239 | 235 | 228 |
| $\Delta C_{el}/\gamma T_c^C$ | 1.16 | 1.32 | 1.13 | 1.34 | 1.35 |

## Figures

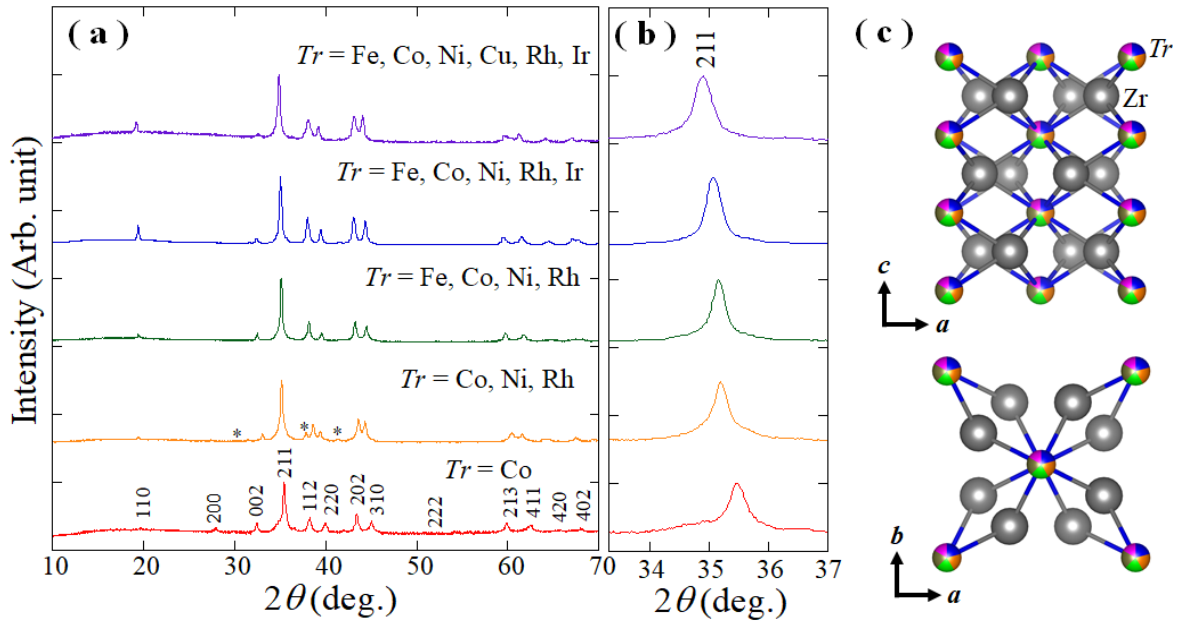

Fig. 1. **(a)** X-ray diffraction (XRD) patterns for the obtained $Tr\text{Zr}_2$ samples (#A) $CoZr_2$, (#B) $Co_{0.3}Ni_{0.4}Rh_{0.3}Zr_2$, (#C) $Fe_{0.2}Co_{0.3}Ni_{0.2}Rh_{0.3}Zr_2$, (#D) $Fe_{0.2}Co_{0.2}Ni_{0.2}Rh_{0.2}Ir_{0.2}Zr_2$, and (#E) $Fe_{0.1}Co_{0.2}Ni_{0.2}Cu_{0.1}Rh_{0.2}Ir_{0.2}Zr_2$. The asterisks indicate an impurity of cubic $CoZr_2$. **(b)** Peak shifts of the 211 peak. **(c)** Schematic images of the crystal structure of $Tr\text{Zr}_2$ with five $Tr$ elements.

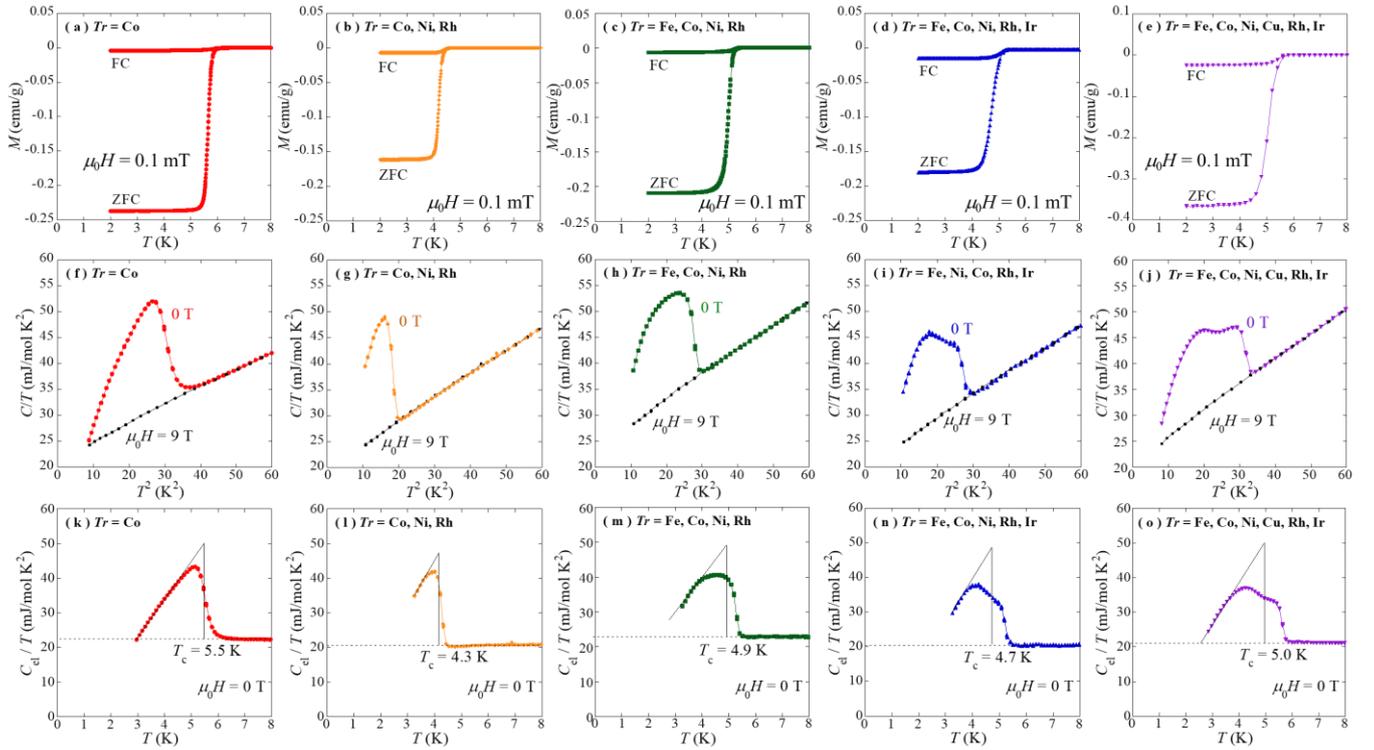

Fig. 2. **(a–e)** Temperature dependences of magnetization $M$ for samples (#A) $CoZr_2$, (#B) $Co_{0.3}Ni_{0.4}Rh_{0.3}Zr_2$, (#C) $Fe_{0.2}Co_{0.3}Ni_{0.2}Rh_{0.3}Zr_2$, (#D) $Fe_{0.2}Co_{0.2}Ni_{0.2}Rh_{0.2}Ir_{0.2}Zr_2$, and (#E) $Fe_{0.1}Co_{0.2}Ni_{0.2}Cu_{0.1}Rh_{0.2}Ir_{0.2}Zr_2$. **(f–j)** Temperature dependence of specific heat $C/T$ for samples #A–#E at 0 and 9 T. **(k–o)** Temperature dependence of electronic specific heat $C_{el}/T$ for #A–#E under 0 T.

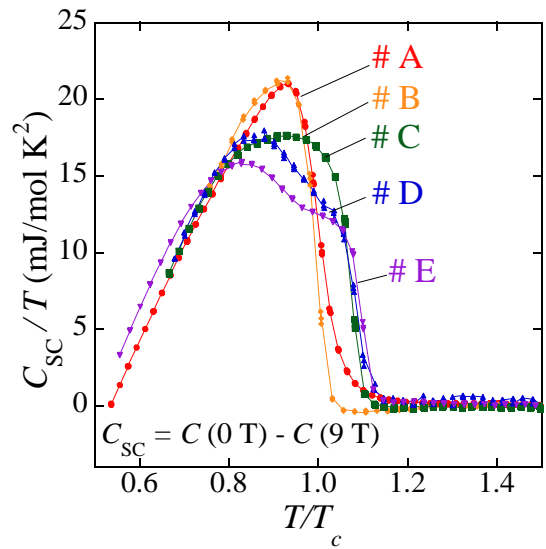

Fig. 3. Comparison of the $C_{sc}/T$ against $T/T_c$ for (#A) $CoZr_2$, (#B) $Co_{0.3}Ni_{0.4}Rh_{0.3}Zr_2$, (#C) $Fe_{0.2}Co_{0.3}Ni_{0.2}Rh_{0.3}Zr_2$, (#D) $Fe_{0.2}Co_{0.2}Ni_{0.2}Rh_{0.2}Ir_{0.2}Zr_2$, and (#E) $Fe_{0.1}Co_{0.2}Ni_{0.2}Cu_{0.1}Rh_{0.2}Ir_{0.2}Zr_2$ under 0 T.

## Supplemental data

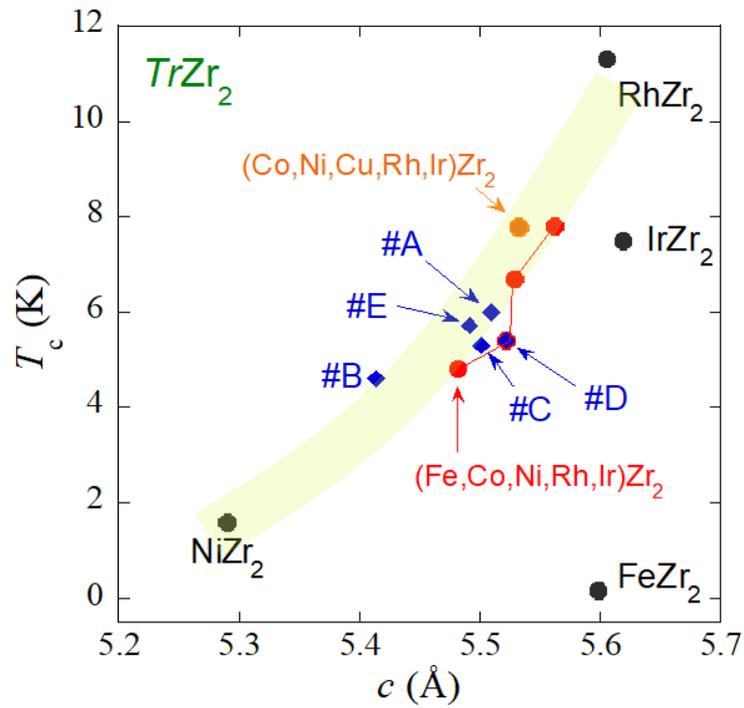

Fig. S1. Lattice constant c dependence of $T_c$ of various $Tr$Zr$_2$ with various $Tr$ elements and $\Delta S_{mix}$. Filled circles indicate datapoints reported in previous works [S1–S3].

[S1] Fisk Z, Viswanathan R, Webb GW 1974 *Solid State Commun.* **15** 1797
[S2] Mizuguchi Y, Kasem MR, Matsuda TD 2021 *Mater. Res. Lett.* **9**, 141
[S3] Kasem MR, Yamashita A, Goto Y, Matsuda TD, Mizuguchi Y 2021 *J. Mater. Sci.* **56** 9499